\documentclass[preprint2]{emulateapj}

\newcommand{\bzcat}{ROMA-BZCAT}

\newcommand{\fer}{{\it Fermi}}

\usepackage{graphicx}
\usepackage{longtable}

\slugcomment{version \today: fm}
\shorttitle{Blazars spectral properties  at 74 MHz}
\shortauthors{F. Massaro et al. 2013}

\begin{document}
\title{Blazars spectral properties  at 74 MHz}
\author{
F. Massaro\altaffilmark{1}, 
M. Giroletti\altaffilmark{2},
A. Paggi\altaffilmark{3}, 
R. D'Abrusco\altaffilmark{3}, 
G. Tosti\altaffilmark{4,5}
\& 
S. Funk\altaffilmark{1}.
}

\altaffiltext{1}{SLAC National Laboratory and Kavli Institute for Particle Astrophysics and Cosmology, 2575 Sand Hill Road, Menlo Park, CA 94025, USA}
\altaffiltext{2}{INAF Istituto di Radioastronomia, via Gobetti 101, 40129, Bologna, Italy}
\altaffiltext{3}{Harvard - Smithsonian Astrophysical Observatory, 60 Garden Street, Cambridge, MA 02138, USA}
\altaffiltext{4}{Dipartimento di Fisica, Universit\`a degli Studi di Perugia, 06123 Perugia, Italy}

\begin{abstract}
Blazars are the most extreme class of active galactic nuclei (AGNs).
Despite a previous investigation at 102 MHz for a small sample of BL Lacs and our recent analysis of blazars
detected in the Westerbork Northern Sky Survey (WENSS), a systematic study of the blazar spectral properties 
at frequencies below 100 MHz has been never carried out.
In this paper, we present the first analysis of the radio spectral behavior of blazars
based on the recent Very Large Array Low-Frequency Sky Survey (VLSS) at 74 MHz.
We search for blazar counterparts in the VLSS catalog confirming that they are detected at 74 MHz.
We then show that blazars present radio flat spectra { (i.e., radio spectral indices $\sim$0.5)} when evaluated also
about an order of magnitude in frequency lower than previous analyses.
Finally, we discuss the implications of our findings in the context of the blazars -- radio galaxies connection
since the low frequency radio data provide a new diagnostic tool to verify the expectations of the unification scenario 
for radio-loud active galaxies.
\end{abstract}

\keywords{galaxies: active - galaxies: BL Lacertae objects -  radiation mechanisms: non-thermal}

\section{Introduction}
\label{sec:intro}
Blazars are compact, core-dominated, radio-loud sources characterized by highly variable, non-thermal, continuum
that extends from radio to $\gamma$-rays. Their spectral energy distributions (SEDs) exhibit two main,
broadly peaked, components: a low-energy one with its maximum between the IR and the X-ray band, 
and the high-energy one peaking in the $\gamma$-rays.
Their emission also features high and variable polarization, 
apparent superluminal motions, and high apparent luminosities \citep[e.g.,][]{blandford78a,urry95}.
Recently, we discovered that their IR colors are clearly distinct
from those of other extragalactic sources, in particular, 
when considering $\gamma$-ray blazars \citep[see also][]{paper1,paper2,paper6}.

Blazars are generally divided in two main categories: the BL Lac objects,
characterized by featureless optical spectra and low luminosity and the 
flat-spectrum radio quasars with optical spectra typical of quasars \\
\citep[see e.g.,][for more details on the optical classification]{stickel91,stoke91,laurent99}.
In the following we label the former class as BZBs and the 
latter as BZQs, adopting the nomenclature 
of the Multifrequency Blazar Catalog\footnote{http://www.asdc.asi.it/bzcat/} \citep[\bzcat,][]{massaro09},
while blazars of uncertain type are indicated as BZUs.

Despite a survey of BZBs at 102 MHz \citep{artyukh81},
the low radio frequency spectral behavior of blazars is still an unexplored portion of the electromagnetic spectrum
for studying the blazar emission.
We recently analyzed blazar data obtained 
with the Westerbork Synthesis Radio Telescope (WSRT),
combining the archival observations present in the 
Westerbork Northern Sky Survey \citep[WENSS;][]{rengelink97} at 325 MHz
with those of the NRAO Very Large Array Sky survey \citep[NVSS;][]{condon98}
and of the Very Large Array Faint Images of the Radio Sky at Twenty-Centimeters \citep[FIRST;][]{becker95,white97} at 1.4 GHz.
We found that blazars have flat radio spectra also between 325 GHz and 1.4 GHz
and on the basis of this result 
we proposed a new approach to search for $\gamma$-ray blazar candidates
among the unidentified gamma-ray sources (UGSs) detected by \fer\ \citep[see][for more details]{paper8}.

In this paper we extend our previous investigation of low radio frequency emission of blazars to frequencies below 100 MHz
using the archival observations of the Very Large Array Low-Frequency Sky 
Survey\footnote{http://lwa.nrl.navy.mil/VLSS/} \citep[VLSS;][]{cohen07}.
The VLSS is a 74 MHz (4m) continuum survey covering the entire sky north of $\sim$-30\degr\ declination. 
The entire survey region is covered with a resolution of $\sim$80\arcsec\ 
with an average root mean square noise of 0.1 Jy/beam. 
We investigate the low-frequency spectral shape of blazars, with particular focus on those that are 
$\gamma$-ray emitters. 
A comparison between this VLSS analysis and that performed using the WENSS is also presented.

In the present work we aim at verifying if blazars maintain a flat radio spectrum even below 100 MHz.
We will also test if their spectral properties are in agreement with the expectations of the unification scenario for 
radio-loud active galaxies which suggests that the observed differences between 
radio galaxies and blazars are mostly due to a different orientation along the line of sight
{ \citep[see][for a recent discussion on radio properties of blazars and radio galaxies in the context of the unification scenario]{kharb10}.}

The paper is organized as follows: in Section~\ref{sec:radius} 
we search for the counterparts of the blazars listed in the \bzcat\ that lie
in area covered by the VLSS while in Section~\ref{sec:sample} we describe the samples used in our analysis.
Section~\ref{sec:blazars} is devoted to description of the low-frequency radio spectral behavior
for the VLSS blazars. Section~\ref{sec:discussion} is dedicated to the discussion of our findings.

For our numerical results,  we use cgs units unless stated otherwise
and we assume a flat cosmology with $H_{\rm 0}=72$ km s$^{-1}$ Mpc$^{-1}$,
$\Omega_{\rm M}=0.26$ and $\Omega_{\Lambda}=0.74$ \citep{dunkley09}.
Spectral indices, $\alpha$, are defined by flux density, S$_{\nu}\propto\nu^{-\alpha}$.

\section{Spatial associations of blazars in the VLSS}
\label{sec:radius}
The starting catalog used in our investigation is the \bzcat\ v4.1, 
listing 3149 blazars \citep[e.g.][]{massaro11}\footnote{www.asdc.asi.it/bzcat/}
distinct as 1220 BZBs (950 BL Lacs and 270 BL Lac candidates), 
1707 BZQs and 222  blazars of uncertain type (BZUs).
However, the \bzcat\ blazars lying above declination $\sim$-30 deg as in the VLSS survey are only 2727: 
1115 BZBs, 1412 BZQs and 200 BZUs. 
In particular, 678 of them are $\gamma$-ray emitters: 349 BZBs, 282 BZQs and 47 BZUs
as associated in the Second \fer\ Large Area Telescope catalog and in the Second \fer\ LAT Catalog of active galactic nuclei 
\citep[2FGL, 2LAC;][respectively]{nolan12,ackermann11a}.

{ The positional uncertainties on the coordinates reported in the \bzcat\ are not uniform since they have been taken from different surveys
but the accuracy on the blazar positions is} of the order of 1\arcsec\ with only several exceptions.
Thus to { identify low radio frequency counterparts of blazars at 74 MHz, we used the following approach, 
already successfully adopted to search for blazar counterparts in the WENSS \citep{paper8}.
We searched for all the \bzcat\ correspondences in the VLSS}
within circular regions of different radii $R$ ranging between 0\arcsec and 30\arcsec. 
To perform our investigation the VLSS catalog available 
on the HEASARC website\footnote{http://heasarc.gsfc.nasa.gov/W3Browse/all/vlss.html} was used.

\begin{figure}[] 
          \includegraphics[height=9.5cm,width=7.2cm,angle=-90]{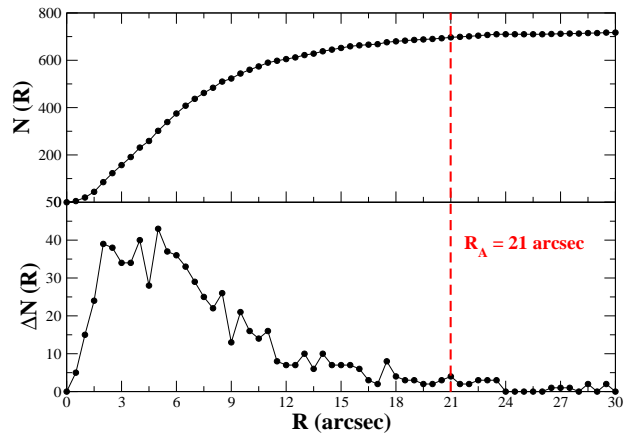}
           \caption{Upper panel) The number of total matches $N(R)$
                        as a function of the radius $R$ between { 0\arcsec\ and 30\arcsec}. 
                        Lower panel) The difference $\Delta N(R)$ between 
                        the number of associations at given radius $R$ and those at $R-\Delta R$
                        as a function of the radius $R$ in the same range of the above plot.
                        The radial threshold $R_A$ selected for our \bzcat\ - VLSS crossmatches is indicated by
                        the vertical dashed red line (see Section~\ref{sec:radius} for more details).}
          \label{fig:radius}
          \end{figure}

We calculated the number of correspondences $N(R)$ as a function of $R$, and
the difference between the number of associations at given radius $R$ and those at ($R-\Delta R$):
          \begin{equation}
          \Delta N(R) = N(R)-N(R-\Delta R)~,
          \end{equation} 
where $\Delta R$= 0.5\arcsec.
Figure~\ref{fig:radius} shows the curves corresponding to $N(R)$ and $\Delta N(R)$ as a function of $R$.
We found that the number of VLSS sources positionally associated with \bzcat\ blazars 
does not increase significantly (i.e., $\Delta N(R)$ systematically lower than 5), at radii larger than 21\arcsec. 
This is also highlighted by the 
correspondent differential curve $\Delta N(R)$, clearly flat for radii greater than 21\arcsec\ 
(see lower panel of Figure~\ref{fig:radius}).
Thus we chose the angular separation of 21\arcsec as the radial association
threshold $R_A$ for assigning VLSS counterparts to \bzcat\ blazars.

The number of spatial associations between the \bzcat\ and the VLSS sources is 697 out of the 2727  
(i.e., $\sim$26\%), all unique matches within $R_A$.
The probability of spurious associations, evaluated by shifting the coordinates 
of the \bzcat\ blazars in random direction of the sky by 1\arcmin, is extremely small being less than 0.1\% 
\citep[see][and references therein for details on the method to estimate the fraction of spurious associations]{maselli10}.

\section{Sample selection}
\label{sec:sample}
We have defined two samples of blazars to carry out our analysis as described below.
{ The main sample, labeled as VLSS Blazar (VLB) sample, 
contains 697 blazars with a unique radio counterpart in the VLSS within 21\arcsec.
while the subsample, labeled as VLSS Gamma-ray Blazar (VLGB)
lists only the { 233} $\gamma$-ray emitting blazars out of 697 sources.}

\begin{figure}[!t] 
          \includegraphics[height=9.5cm,width=7.cm,angle=-90]{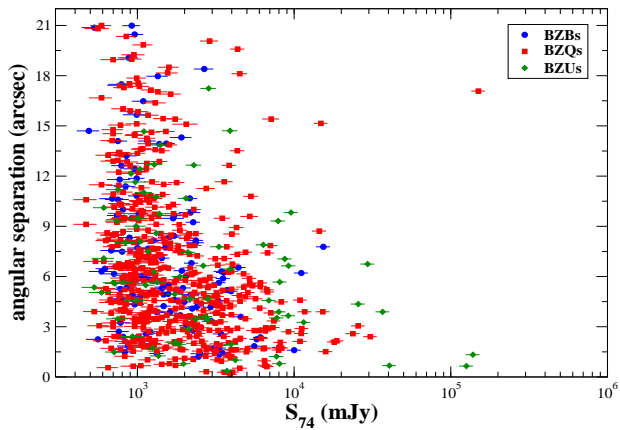}
           \caption{The angular separation between the \bzcat\ positions and that of the VLSS catalog
                         versus the flux density at 74 MHz, $S_{74}$, for the whole VLB sample.}
          \label{fig:angular}
          \end{figure}

In Figure~\ref{fig:angular} we show the scatter plot of the angular separation 
between the \bzcat\ positions and that of the VLSS catalog
as a function of the integrated flux density, $S_{74}$, for the whole VLB sample.
As expected, there is a mild trend between these two quantities because the position of faint VLSS 
sources are less accurately determined \citep[e.g.,][]{cohen07}. 
{ In addition, a larger population of weak sources against 
which the cross-identification could fail to identify the correct counterpart might be also present.}
{ All the blazar samples previously described are summarized in Table~\ref{tab:sample}.}
\begin{table}
\caption{Blazar samples.}
\begin{tabular}{|lcccc|}
\hline
Sample                          & BZBs & BZQs & BZUs & Total \\
\hline
\noalign{\smallskip}
Blazars at Dec.$>$-30\degr\       & 1115 & 1412 & 200  & 2727  \\
VLB sample                      &  128 &  495 &  74  &  697  \\
\noalign{\smallskip}
Fermi blazars at Dec.$>$-30\degr\ &  349 &  282 &  47  &  678  \\ 
VLGB sample                     &   67 &  145 &  21  &  233  \\
\hline
\end{tabular}\\
\label{tab:sample}
\end{table}

\section{Blazar spectral properties at 74 MHz}
\label{sec:blazars}

\subsection{Spectral shape at 74 MHz}
\label{sec:spectra}
The \bzcat\ blazars are associated with NVSS or  FIRST counterparts
so their flux density at 1.4 GHz: $S_{1400}$ is available in the catalog.
In particular, for 692 out the 697 in the VLB sample the flux at 4.85 GHz is also reported in the \bzcat.
Thus we { calculated the following two radio spectral indices:
$\alpha_{74}^{1400}$ between 74 MHz and 1.4 GHz and that $\alpha_{1400}^{4850}$ between 1.4 GHz and 4.85 GHz}
and we also computed the difference:
\begin{equation}
\Delta\alpha = \alpha_{1400}^{4850} - \alpha_{74}^{1400}~.
\end{equation}
Thus, sources with $\Delta\alpha<$0 show radio spectral shapes that steepen moving toward low frequencies
while those with  $\Delta\alpha>$0 appear to be flatter below 1.4 GHz than they are at high radio frequencies.
{ The relative uncertainties on the radio flux densities are $\sim$13\%, 3\% and 9\% at 74 MHz, 1.4 GHz and 4.85 GHz respectively,
in particular being systematically lower than 10
This corresponds to a mean uncertainty on the spectral indices of 0.04 and 0.08 for $\alpha_{74}^{1400}$ and $\alpha_{1400}^{4850}$, respectively
and to a mean error on $\Delta\alpha$ of 0.09\footnote{For $\alpha=log(S_2/S_1)/log(\nu_2/\nu_1)$ 
the correspondent error is $\sigma_\alpha=1/log(\nu_2/\nu_1)/ln(10) \cdot \sqrt{\rho_1^2+\rho_2^2}$ where 
$\rho_i=\sigma_i/S_i$ (with i=1,2) is the relative error on each flux density.}.
Thus the uncertainties on the spectral indices are definitively not affecting our analysis since the standard deviation of the distributions
for the entire VLB sample is systematically larger than them.}

\begin{figure}[!t] 
          \includegraphics[height=9.5cm,width=6.5cm,angle=-90]{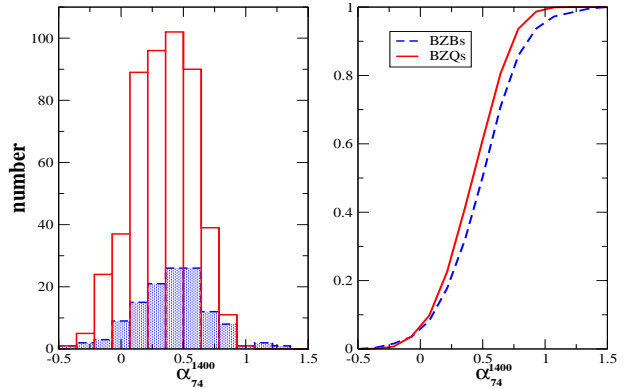}
           \caption{Left panel) The distributions of the radio spectral index $\alpha_{74}^{1400}$
                         for blazars in the VLB sample, BZBs (blue) and BZQs (red).
                         Right panel) The cumulative distribution for the same sample.}
          \label{fig:hist_bzcat}
          \end{figure}

In Figure \ref{fig:hist_bzcat}, we show the distribution of the $\alpha_{74}^{1400}$
for the blazars in the VLB sample distinguishing between the BZBs and BZQs.
The large fraction of radio spectral indices, $\alpha_{74}^{1400}$, for blazars are systematically smaller than 1.0 
(i.e., $\sim$99\%), with the 80\% lower than $\sim$0.75. 
In particular, the two $\alpha_{74}^{1400}$ { distributions} for the BZBs and the BZQs
appear to be similar at 99\% level of confidence according to a Kolmogorov-Smirnov (KS) test. 
These { distributions indicate} that they have relatively flat radio spectra
also at frequencies below 100MHz, in agreement with their behavior at 325 MHz \citep{paper8}.
The flatness of the blazar radio spectra is expected from the  
high radio frequency data in the GHz energy range \citep[e.g.,][]{healey07,ivezic02,kimball08} 
and this spectral property was also used for the identification of $\gamma$-ray sources
since the EGRET era \citep[e.g.,][]{mattox97}.
However, low-frequency radio observations, as those of the VLSS at 74 MHz 
were never previously investigated to confirm this trend for a systematic study of the blazar population.
          \begin{figure}[!t] 
          \includegraphics[height=9.5cm,width=6.5cm,angle=-90]{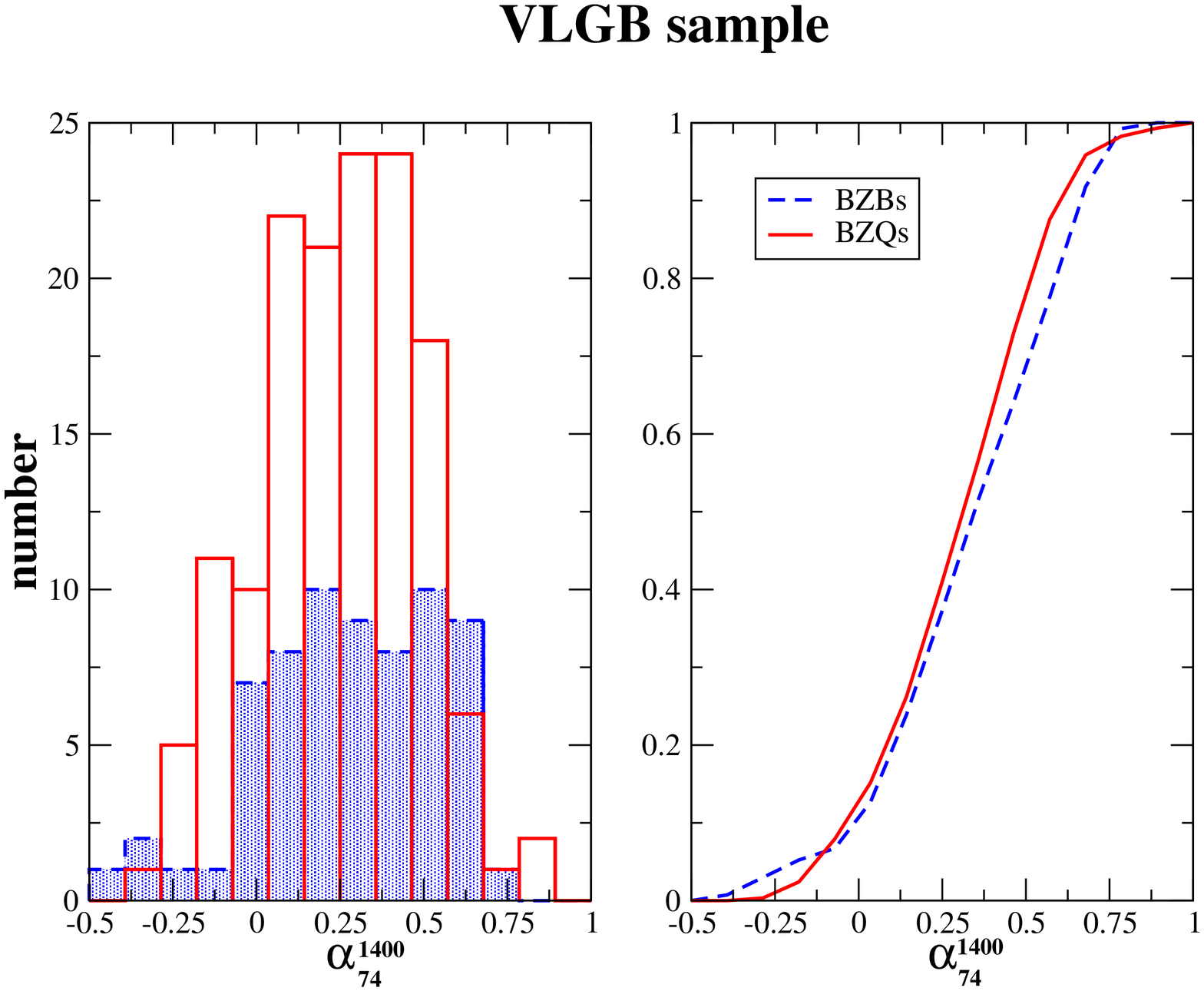}
          \caption{Left panel) The distribution of the radio spectral index $\alpha_{325}^{1400}$
                       of all the blazars in the VLGB sample, BZBs (blue) and BZQs (red).
                       Right panel) The cumulative distribution for the same sample.}
          \label{fig:hist_fermi}
          \end{figure} 

In Figure~\ref{fig:hist_fermi}, we show the comparison between the BZB and the BZQ 
spectral index distributions considering only $\gamma$-ray blazars in the VLGB sample.
Here we note that all the \fer\ detected blazars have spectral indices systematically flatter than
the those in the VLB sample. 
As observed in the VLB sample (see Figure~\ref{fig:hist_bzcat}), 
a KS test indicates that also in this case these two $\alpha_{74}^{1400}$ distributions of BZBs and BZQs in the VLGB sample
are similar at 99\% level of confidence.
In addition, we did not find any correlation or net trend between the $\alpha_{74}^{1400}$ and the $\gamma$-ray 
spectral index $\alpha_{\gamma}$ for the whole VLGB sample.

\subsection{Flux densities at 74 MHz}
\label{sec:fluxes}
\begin{figure}[] 
          \includegraphics[height=9.5cm,width=6.5cm,angle=-90]{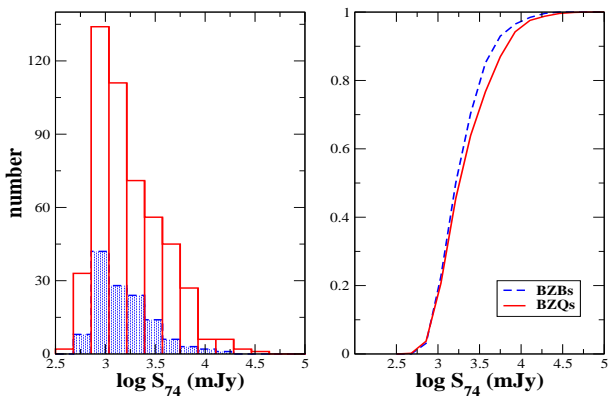}
           \caption{Left panel) The distributions of the low-frequency radio flux density $S_{74}$
                         for the blazars in the VLB sample.
                         The BZB and BZQ distributions appear statistically similar according to the KS test.
                         Right panel) The cumulative distribution for the same sample.}
          \label{fig:hist_flux}
          \end{figure}

\begin{figure}[!b] 
          \includegraphics[height=9.5cm,width=6.5cm,angle=-90]{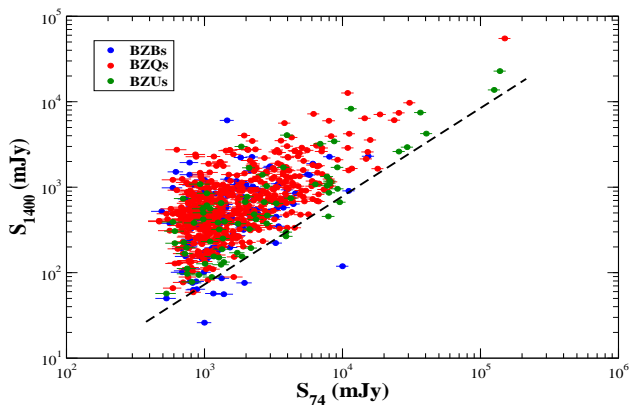}
           \caption{The scatter plot between the $S_{1400}$ and $S_{74}$ radio flux densities.
                         The BZBs (blue circles), BZQs (red square) and BZUs (green diamonds) are shown separately.
                         The black dashed line corresponds to the power-law: $S_{1400}$=0.05$\times S_{74}$.}
          \label{fig:fluxesradio}
          \end{figure}

We computed the distributions of the $S_{74}$ flux densities comparing the BZBs and the BZQs
for the VLB sample as shown in Figure~\ref{fig:hist_flux}, that appear similar
at 99\% level of confidence evaluated according to a KS test.
In Figure~\ref{fig:fluxesradio} we show the scatter plot of the NVSS flux density at 1.4 GHz $S_{1400}$ with respect to
that at 74 MHz $S_{74}$ where it is worth noting that 99\% of the blazars detected in the VLB sample
have { $S_{1400}$=0.05$\times S_{74}$, as indicated by the black dashed line.}

\begin{figure}[] 
          \includegraphics[height=9.5cm,width=7.cm,angle=-90]{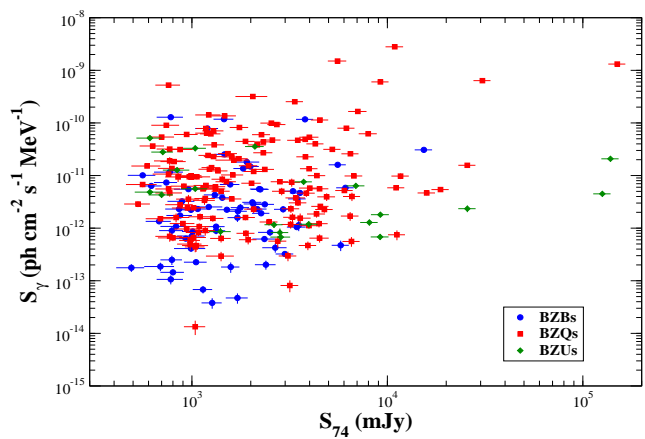}
           \caption{The scatter plot of the $\gamma$-ray flux density $vs$ radio 74 MHz 
                        for the BZBs (blue), the BZQs (red) and the BZUs (green)
                        that belong to the VLGB sample (see Section~\ref{sec:fluxes} for more details).}
          \label{fig:fluxesradiogamma}
          \end{figure}

For the blazars in the VLGB sample, we also searched for a trend between the radio and the $\gamma$-ray emissions 
\citep[e.g.,][]{ghirlanda10,mahony10,ackermann11b} as we did previously using the WENSS observations at 325 MHz \citep{paper8};
however, as shown in Figure~\ref{fig:fluxesradiogamma} there is not a clear trend or
correlation between the $\gamma$-ray and the radio flux density at 74 MHz.

We remark that only 697 out of 2727 \bzcat\ sources have a unique association in the VLSS survey
within a radius of 21\arcsec, thus $\sim$74\% of them were { not} associated or either not detected at 74 MHz.
Most of these undetected blazars could lie below the completeness threshold of the VLSS, that for the 50\% point-source 
detection limit is roughly 0.7 Jy for a typical noise level of 0.1 Jy/beam,
evaluated taking into account both ionospheric smearing and the clean bias \citep{cohen07}.
In addition, the noise levels are not constant throughout the survey region and only the 
differential completeness as a function of signal-to-noise ratio 
has been defined \citep[see Section 8 of][for more details]{cohen07}.
Consequently, it is not possible to extrapolate the $S_{1400}$ flux densities to 74 MHz to test if blazars 
with no VLSS counterparts lie above or below the completeness of the survey.
This situation is even more complicated because blazars could be variable
at 74 MHz making any expectation less reliable.

However, to evaluate if we should expect to detect the remaining 2030 blazars,
we assigned a spectral index $\alpha_{74}^{1400}$ equal to the peak of the spectra index distributions 
of the BZBs, BZQs and BZUs in the VLB sample to those blazars that do not have a VLSS counterpart
within 21\arcsec\ and we computed the extrapolated $S^*_{74}$ flux density on the basis of their measured $S_{1400}$.
We adopted a value of $\alpha_{74}^{1400}$ equal to 0.43, 0.34 and 0.51 for the BZBs, the BZQs and the BZUs, respectively.
Comparing the VLSS radio sources with the blazars in the VLB sample
we found that 63\% of the VLSS objects have flux densities $S_{74}$ greater than $\sim$ 1Jy
while only $\sim$16\% of the undetected blazars have extrapolated $S^*_{74}$ above this threshold,
indicating that the large fraction of them are not expected to have a counterpart in the VLSS as occurs.

\section{Discussion and conclusions}
\label{sec:discussion}
We investigated the distribution of the radio spectral index $\alpha_{74}^{1400}$ 
for the blazars with a VLSS counterpart also focusing on those that are associated with $\gamma$-ray sources.
We found that about 60\% of $\gamma$-ray emitting blazars have flat radio spectra (i.e., $\alpha_{74}^{1400}<$0.5),
with 99\% even smaller than 0.9 (see Section~\ref{sec:spectra} for more details)
as occurs when analyzing radio data at higher frequency \citep[e.g.,][]{healey07,paper8}.
This strongly suggests that blazar spectra are still dominated by the beamed radiation arising from 
particles accelerated in their relativistic jets even at 74 MHz, 
in agreement with the recent results of Kimball et al. (2011).
          \begin{figure}[] 
          \includegraphics[height=9.2cm,width=7.cm,angle=-90]{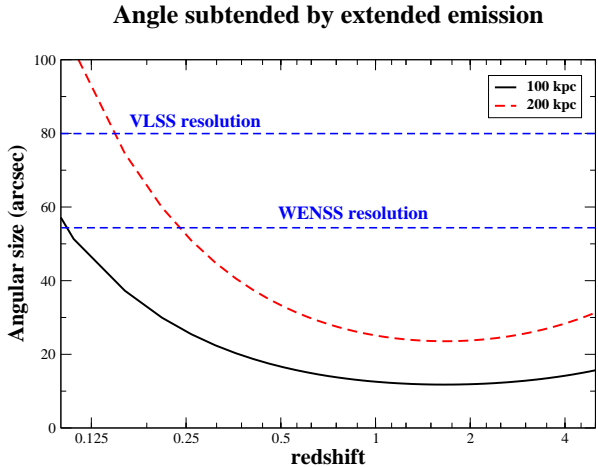}
           \caption{The angle subtended by extended structures of 100 kpc (black line) and 200 kpc (red dashed line), 
                        respectively as a function of the redshift. { The two horizontal blue 
                        dashed lines} represent the typical angular resolution 
                        of the WENSS\citep{rengelink97} and VLSS \citep{cohen07} surveys.
                        It is clear that for blazars at redshifts larger than $\sim$0.1, 
                        extended structures of size smaller than 200 kpc
                        cannot be resolved. This indicates that the flux densities measured at low radio frequencies
                        arise from both nuclear and large scale components.
                        The 100 kpc and 200 kpc size are representative of the scale observed in radio galaxies.}
          \label{fig:cosmo}
          \end{figure}

There are several implications of our results in the context of the unification scenario of radio-loud AGNs.
In 1974 Fanaroff and Riley proposed the classification scheme
for the extragalactic radio sources distinguishing two classes on the basis of the correlation between the relative
positions of regions of high and low surface brightness in their extended components \citep{fanaroff74}.
They introduced the ratio $R_{FR}$ of the distance between the regions of highest surface
brightness on opposite sides of the central galaxy and/or quasar, to
the total extent of the source up to the lowest brightness contour in the radio map.  
Sources with $R_{FR}$ $\leq$ 0.5 were placed in Class I (i.e., FR\,I) 
and sources with $R_{FR}$ $\geq$ 0.5 in Class II
(i.e., FR\,II).  At radio frequencies, FR\,Is show surface brightness higher 
toward their cores while FR\,IIs toward their edges.
It was also found that nearly all sources with luminosity
$L_{178 MHz}$ $\leq$  2 $\times$ 10$^{25}$ $h_{100}^{-2}$ W Hz$^{-1}$ str$^{-1}$
were FR\,I while the brighter sources were nearly all of FR\,II. 
The luminosity boundary between them is not very sharp, and there is some overlap 
in the luminosities of sources classified as FR\,I or FR\,II on the basis of their structures.

Several observational evidences support the idea originally proposed by Blanford \& K$\ddot{o}$nigl in 1979
that suggested powerful FR\,II radio galaxies as the parent populations of BZQs while 
BZBs were assumed intrinsically similar to weak FR\,Is
\citep[e.g.,][ and reference therein for more details]{urry95,scarpa01}. 
Both radio galaxies and blazars have similar host galaxies: giant ellipticals \citep[e.g.,][]{scarpa00a,scarpa00b}
and deep radio observations of selected blazars at 1.4 GHz show extended structures 
remarkably similar to those of lobes and plumes of radio galaxies \citep[e.g.][]{antonucci85}.
Thus, according to the unification scenario of radio-loud active galaxies\\ \citep[e.g.,][]{blandford78b,blandford79,urry95},\\
radio galaxies are generally interpreted as misaligned blazars.

The low-frequency radio observations of blazars presented here 
allow us to make a direct test of the unification scenario of radio-loud active galaxies
and of the blazars -- radio galaxies connection. 
At MHz frequencies radio galaxies clearly show steep radio spectra 
as for example highlighted in the Third Cambridge Catalog of radio sources \citep[3C;][]{edge59,spinrad85}
by Kellerman et al. (1968) and Pauliny-Toth et al. (1968) \citep[see also][]{kellerman69a,kellerman69b}.
The main reason underlying this spectral behavior 
is due to the combination of emission arising from compact cores, having typically flat radio spectra,
with that of extended structures, such as plumes (in FR\,Is) or lobes (in FR\,IIs) 
characterized by steep radio spectra. Consequently, their combined spectra at low-frequency, such as at 74 MHz,
where the large beam of the radio surveys does not allow us to resolve different components, 
is dominated by that of large (i.e., kpc) scale structures thus resulting in steep radio spectra. 
In particular, Figure~\ref{fig:cosmo} shows the angle subtended by 100 kpc and 200 kpc extended structures
as a function of redshift $z$. It is clear that above $z \sim$0.1, both the VLSS and the WENSS surveys cannot 
distinguish the emission arising { from} the core and from the extended components in radio galaxies and blazars.
Thus flux densities measured at 74 MHz include both these contributions.

\begin{figure}[] 
          \includegraphics[height=9.2cm,width=7.cm,angle=-90]{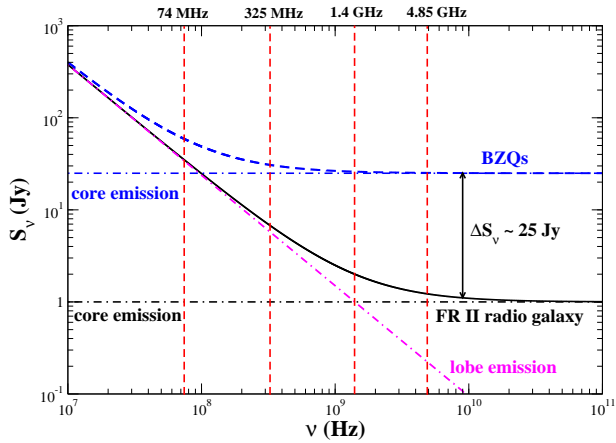}
           \caption{{\it Left panel)} The schematic view of the integrated radio spectrum of an FR\,II radio galaxy
                        { (black curve)} in comparison with that of a BZQ (blue curve),
                        assuming the same emission arising from their extended structures. 
                        The flat spectrum core emission showing dominates at high frequencies 
                        while the steep spectrum of the extended structures (i.e., lobes) 
                        overcomes the nuclear radio at low frequencies.
                        However, since BZQs show relatively flat spectra { at} low radio frequencies
                        the core emission has to be much brighter than that of the FR\,II radio galaxy, 
                        as for example, up to a factor of $\sim$25 when compared at $\sim$5 GHz. 
                        This scenario is in agreement with the expectations of the
                        unification scenario of radio-loud active galaxies.}
          \label{fig:spectrum}
          \end{figure}

{ Since the emission arising from large scale structures
is isotropic, it is not dependent on the orientation relative to the line of sight.
So, according to the unification scenario,
we expect to detect steep radio spectra, typical of extended components, also when observing blazars
at low radio frequencies as occurs in radio galaxies.
To describe the above situation we can consider the schematic representation of the radio spectra
of a BZQ in comparison with that of an FR\,II radio galaxy as shown in Figure~\ref{fig:spectrum}.
Given the resolution of the low frequency radio surveys (e.g., Figure~\ref{fig:cosmo}) 
that cannot resolve and separate the contributions of extended and nuclear components, 
the integrated spectrum of a BZQ is expected to be similar to that of an FR\,II radio galaxy at low frequencies,
unless the core emission is overcoming that of the extended structures (i.e., $\Delta\alpha>0$).}

\begin{figure}[!b] 
          \includegraphics[height=9.2cm,width=7.cm,angle=-90]{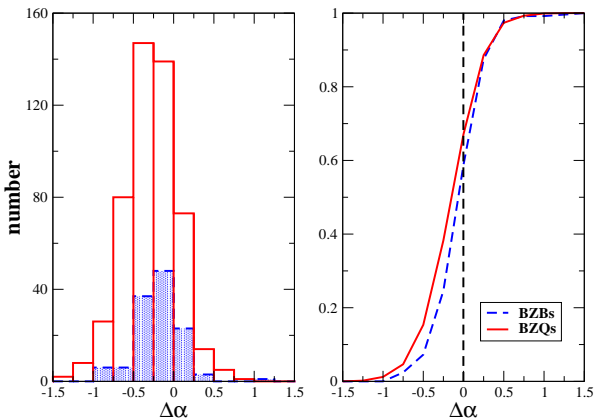}
           \caption{The distribution of the $\Delta\alpha = \alpha_{1400}^{4850} - \alpha_{74}^{1400}$ for the BZBs (blue) and the BZQs (red)
           in the VLB sample for which radio data at 4.85 GHz are available (i.e., 692 out of 697 sources).
           The cumulative distribution is also shown on the right panel. The dashed vertical black line corresponds to the dividing sign
           between blazars that become steeper ($\Delta\alpha<0$) or flatter ($\Delta\alpha>0$) at low frequencies
           that they are above 1.4 GHz.}
          \label{fig:hist_delta}
          \end{figure}

\begin{figure}[] 
          \includegraphics[height=9.2cm,width=7.cm,angle=-90]{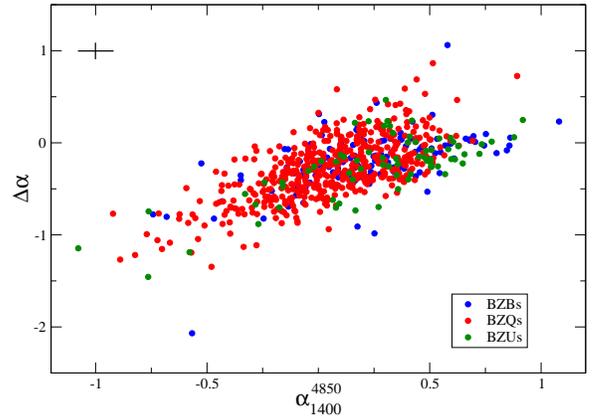}
           \caption{The scatter plot of the spectral index variation $\Delta\alpha$ as a function of $\alpha_{1400}^{4850}$
           for BZBs (blue circles), BZQs (red squares) and BZUs (green diamond), respectively. 
           Blazars with flatter radio spectra tend to have steep components dominating at low frequencies below 1.4 GHz.
           { The black cross indicates the typical uncertainty on the $\Delta\alpha$ and $\alpha_{1400}^{4850}$ values.}}
          \label{fig:delta-alpha}
          \end{figure}

\begin{figure}[] 
          \includegraphics[height=9.2cm,width=7.cm,angle=-90]{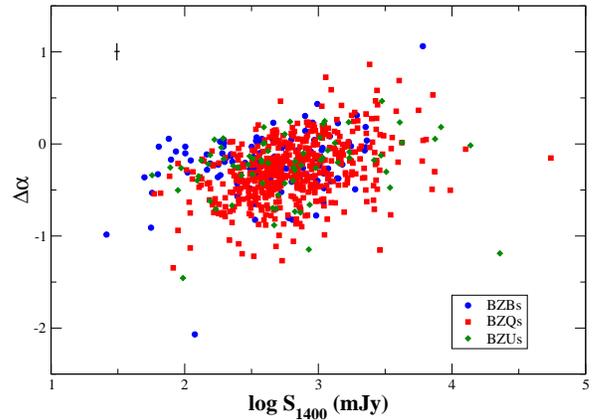}
           \caption{The scatter plot of the spectral index variation $\Delta\alpha$ as a function of the radio flux density at 1.4 GHz $S_{1400}$
           for BZBs (blue circles), BZQs (red squares) and BZUs (green diamond), respectively. 
           Blazars bright at 1.4 GHz tend to hide the emission arising from extended structures { (i.e., steep components)} 
           at low frequencies below 1.4 GHz.
           { The black cross indicates the typical uncertainty on the $\Delta\alpha$ and $S_{1400}$ values.}}
          \label{fig:delta-flux}
          \end{figure}

In order to test these expectations of the unification scenario we computed the $\Delta\alpha$ 
for all the sources in the VLB sample that have data at 4.85 GHz as reported in the \bzcat.
In Figure~\ref{fig:hist_delta}, we show the distribution of the $\Delta\alpha$ for the BZBs and the BZQs; 
it is quite evident that the large fraction of the VLB sources (i.e., $\sim$70\%) show the spectral shape
expected by the interpretation of blazars being intrinsically similar to radio galaxies. 
In addition, we plot the $\Delta\alpha$ as a function of the radio spectral index evaluated at high frequencies $\alpha_{1400}^{4850}$
(Figure~\ref{fig:delta-alpha}) and of the radio flux density at 1.4 GHz $S_{1400}$ (Figure~\ref{fig:delta-flux}) .
These two additional plots also show trends in agreement with the unification scenario of radio-loud sources, since 
radio emission from steep components tend to be more evident in blazars with flatter or even inverted radio spectra 
(see also Figure~\ref{fig:delta-alpha}) and brighter sources tend to maintain their flat spectra also at low frequencies,
according to the fact that their nuclear beamed radiation overwhelms that arising from their extended structures.
However, one controversy arises from the above results.
It is indeed clear the presence of blazars with counterparts for which the $\Delta\alpha>0$.
This spectral difference could be interpreted as, for example, 
due to synchrotron self absorbed components in the blazar compact cores, 
or to intrinsic source variability of the epochs when the radio observations were taken,
since they are not simultaneous.

{ In particular, blazars show intrinsic variability over a wide range of wavelengths 
and their flux variations, even at radio wavelengths, occur stochastically, so this could affect the estimate of the spectral indices.}
However, the VLB sample is selected on the basis of the source detection at 74 MHz implying that
these blazars have 1.4 GHz and 4.85 GHz flux densities well above the sensitivity limit of the high-frequency radio surveys.
{ We expect that the flux densities measured at 1.4 GHz and 4.85 GHz are well representative 
and thus the spectral indices between these two frequencies are also well representative of the average state of the source.}
In the X-ray band, this generic survey property has been extensively used to select blazars as targets
for investigating the absorption lines in the warm hot intergalactic region \citep[see e.g.,][]{nicastro03,nicastro05,nicastro13}.

Consequently, given the absence of any duty cycle in the blazar variability pattern and the random occurrence of their flaring activity,
for each VLB source having a $\Delta\alpha>0$, due, for example, to a high state { occurring} while measuring the flux density at 1.4 GHz,
we could expect a blazar with $\Delta\alpha<0$ due to a flare at 74MHz while { the VLSS observations were performed.}

In addition, the scatter induced on the $\Delta\alpha$ distribution remains below $\sim 0.17$,
i.e., half as small as the standard deviation of the real distribution ($\sim 0.33$, see Figure~\ref{fig:hist_delta}), 
for flux variations up to a factor of 1.5. 
This value is larger than the typical variations observed for most blazars 
over the course of a dedicated multi-year monitoring at 15 GHz, 
where the flux changes are expected to be even larger than those at 1.4 GHz or below \citep{richards11}.

Thus the presence of VLB sources with $\Delta\alpha>0$, if not affected by variability,
is not completely in agreement with the unification scenario of radio-loud active galaxies 
since it would imply the presence of an appreciable fraction of blazars 
for which radio emission from extended structures (plumes and/or lobes) is not detected (as indeed expected).

This discovery could imply new insights on the unification scenario of radio-loud active galaxies
opening new questions as for example:
i) is the distinction between blazars and radio galaxies driven not only by the orientation along the line of sights
but also by a different parameter or a combination of them? (e.g., accretion rate, black hole mass)
ii) Is it possible that the difference between radio-loud active galaxies could be also due to their surrounding environments?
{ The unsolved issues deserve further investigation to confirm the peculiar spectral behavior of 
blazars. Independent analysis can be carried out using the 
low frequency observations of the Murchison Widefield Array (MWA) \citep{tingay13} radio telescope combined with 
existing Australia Telescope Compact Array (ATCA) surveys at 20 GHz \citep[e.g.,][]{frater92,murphy10,massardi11}.
Future investigations with the new generation of low-frequency radio telescopes 
as the LOw Frequency ARray (LOFAR) \citep[e.g.,][]{vanhaarlem13}, the Square Kilometer Array (SKA) \citep[e.g.,][]{dewdney10}
will also be crucial to resolve nuclear and extended components in blazars.}

\acknowledgements
{ We thank our anonymous referee for many helpful comments and for stimulating the discussion on the 
possible effects of the variability, which greatly improved this manuscript.}
We thank R. Morganti and D. Harris for their valuable comments and suggestions.
F. Massaro is grateful to S. Digel, E. Mahony, M. Murgia, Howard Smith and M. Urry for their helpful discussions.
The work is supported by the NASA grants NNX12AO97G.
R. D'Abrusco gratefully acknowledges the financial 
support of the US Virtual Astronomical Observatory, which is sponsored by the
National Science Foundation and the National Aeronautics and Space Administration.
The work by G. Tosti is supported by the ASI/INAF contract I/005/12/0.
The WENSS project was a collaboration between the Netherlands Foundation 
for Research in Astronomy and the Leiden Observatory. 
We acknowledge the WENSS team consisted of Ger de Bruyn, Yuan Tang, 
Roeland Rengelink, George Miley, Huub Rottgering, Malcolm Bremer, 
Martin Bremer, Wim Brouw, Ernst Raimond and David Fullagar 
for the extensive work aimed at producing the WENSS catalog.
Part of this work is based on archival data, software or on-line services provided by the ASI Science Data Center.
This research has made use of data obtained from the High Energy Astrophysics Science Archive
Research Center (HEASARC) provided by NASA's Goddard
Space Flight Center; the SIMBAD database operated at CDS,
Strasbourg, France; the NASA/IPAC Extragalactic Database
(NED) operated by the Jet Propulsion Laboratory, California
Institute of Technology, under contract with the National Aeronautics and Space Administration.
Part of this work is based on the NVSS (NRAO VLA Sky Survey)
and on the VLA low-frequency Sky Survey (VLSS);
The National Radio Astronomy Observatory is operated by Associated Universities,
Inc., under contract with the National Science Foundation. 
This publication makes use of data products from the Two Micron All Sky Survey, which is a joint project of the University of 
Massachusetts and the Infrared Processing and Analysis Center/California Institute of Technology, funded by the National Aeronautics 
and Space Administration and the National Science Foundation.
This publication makes use of data products from the Wide-field Infrared Survey Explorer, 
which is a joint project of the University of California, Los Angeles, and 
the Jet Propulsion Laboratory/California Institute of Technology, 
funded by the National Aeronautics and Space Administration.
TOPCAT\footnote{\underline{http://www.star.bris.ac.uk/$\sim$mbt/topcat/}} 
\citep{taylor2005} for the preparation and manipulation of the tabular data and the images.

{Facilities:} \facility{VLA}, \facility{WSRT}

{}

\end{document}